\documentclass{aa} 
\usepackage{graphicx} 

\begin{document}

   \title{On the origin of nitrogen}

\author{L.S.~Pilyugin \inst{1},  
        T.X.~Thuan  \inst{2},
        J.M.~V\'{\i}lchez \inst{3} }

  \offprints{L.S. Pilyugin }

   \institute{   Main Astronomical Observatory
                 of National Academy of Sciences of Ukraine,
                 27 Zabolotnogo str., 03680 Kiev, Ukraine, 
                 (pilyugin@mao.kiev.ua)
                 \and
                 Astronomy Department, University of Virginia, Charlottesville, VA 22903, 
                 (txt@virginia.edu)
                 \and
                 Instituto de Astrof\'{\i}sica de Andaluc\'{\i}a,
                 Apdo, 3004, 18080 Granada, Spain  
                 (jvm@iaa.es)
                 }
                 
\date{Received 30 May 2002 / accepted 00  Month 2002}

\abstract{
The problem of the origin of nitrogen is considered within the framework of 
an empirical approach.  The oxygen abundances and nitrogen to oxygen 
abundances ratios are derived in H\,{\sc ii} regions of a number of spiral 
galaxies through the recently suggested P -- method using more than six hundred 
published spectra. The N/O -- O/H diagram for H\,{\sc ii} regions in 
irregular and spiral galaxies is constructed. It is found that the N/O values 
in H\,{\sc ii} regions of spiral galaxies of early morphological types 
are higher than those in H\,{\sc ii} regions with the same metallicity in 
spiral galaxies of late morphological types. This suggests a long-time-delayed 
contribution to the nitrogen production. The N/O ratio of a galaxy can then be 
used as an indicator of the time that has elapsed since the bulk of star 
formation occurred, or in other words of the nominal "age" of the galaxy 
as suggested by Edmunds \& Pagel more than twenty years ago. The scatter in N/O 
values at a given O/H can be naturally explained by differences in star 
formation histories in galaxies. While low-metallicity dwarf galaxies with low 
N/O do not contain an appreciable amount of old stars, low-metallicity dwarf 
galaxies with an appreciable fraction of old stars have high N/O. 
Consideration of planetary nebulae in the Small Magellanic Cloud 
and in the Milky Way Galaxy suggests that the contribution of low-mass stars to the 
nitrogen production is significant, confirming the conclusion that there 
is a long-time-delayed contribution to the nitrogen production. 
}             

\titlerunning{On the origin of nitrogen}

\authorrunning{L.S.~Pilyugin, T.X.~Thuan, J.M.~ V\'{\i}lchez}  

\maketitle

\section{Introduction}

The origin of nitrogen has always been one of the central problems of the 
theory of chemical evolution of galaxies. 
The theoretical stellar yields of nitrogen have been computed by many 
investigators (Renzini \& Voli 1981; Maeder 1992; Marigo et al. 1996, 1998, 
2001; van den Hoek \& Groenewegen 1997; Boothroyd \& Sackmann 1999; 
Rauscher et al. 2002; Meynet \& Maeder 2002a,b; among others).  
However, the results are not undisputed. 
Therefore, the "empirical" approach to study the origin of nitrogen, i.e     
the comparison of the predictions of models of chemical evolution of 
galaxies computed with different assumptions for nitrogen yields with 
observed N abundances, has been widely used
(Edmunds \& Pagel 1978; Lequeux et al. 1979; Serrano \& Peimbert 1983; 
Matteucci \& Tosi 1985; Matteucci 1986; Garnett 1990; Pilyugin 1992, 1993, 1999;  
Vila-Costas \& Edmunds 1993; Marconi et al. 1994; Thuan et al. 1995; Thurston 
et al. 1996; Kobulnicky \& Skillman 1996, 1998; van Zee et al. 1998a; Izotov \& 
Thuan 1999; Coziol et al. 1999; Henry et al. 2000; Contini et al. 2002; among others).
The interpretation of the observed N/O vs. O/H  diagram is at the base of 
the empirical approach. 

Thanks to the work of Thuan et al. (1995), Kobulnicky \& Skillman (1996),
van Zee et al. (1997), Izotov \& Thuan (1999), and others, the N/O vs. O/H diagram 
appears to be well established in the low-metallicity range. 
In the low-metallicity range, it has significantly changed during the last decade. There is no 
commonly accepted interpretation of the present-day N/O vs. O/H  diagram.
Izotov \& Thuan (1999) have concluded that massive stars are responsible 
for the nitrogen abundances observed in low-metallicity blue compact galaxies, 
while Henry et al. (2000) have found that intermediate-mass stars make a 
dominant contribution to the nitrogen production. 
As noted by Henry et al. (2000), a major problem with N/O ratios has been 
to try to explain the spread in N/O at a given O/H. Close examination of 
the suggested explanations of the origin of the scatter in the present-day N/O 
-- O/H diagram shows that existing explanations are not  
satisfactory (see below). This provides the motivation for this study.

We use also here an empirical approach to study the origin of nitrogen. 
Why should we hope that this attempt would be more successful than previous 
ones based on the empirical approach in establishing the origin of nitrogen? 
Most studies on the origin of nitrogen have concentrated on low-luminosity 
dwarf galaxies. Van Zee et al. (1998a) have shown that the use of 
low-metallicity H\,{\sc ii} regions at the periphery of spiral galaxies for 
investigation of the origin of nitrogen provides some advantages in comparison 
to H\,{\sc ii} regions in dwarf irregular galaxies. In previous studies, the 
oxygen abundances in high-metallicity H\,{\sc ii} regions, where 
[OIII]$\lambda 4363$ line is usually not available, were derived using 
an empirical method (usually the R$_{23}$ -- method suggested by Pagel et al. 
1979). Recently it has been shown (Pilyugin 2000, 2001a,b) that the oxygen 
abundance derived with the R$_{23}$ -- method produces a systematic error
depending on the excitation parameter P: the R$_{23}$ -- method provides more
or less realistic oxygen abundances in high-excitation H\,{\sc ii} regions, but
yields overestimated oxygen abundances in low-excitation H\,{\sc ii} regions.
This is in agreement with the result of Kinkel \& Rosa
(1994), who showed the need of lowering all H\,{\sc ii} region abundances
obtained with the R$_{23}$ calibration of Edmunds and Pagel (1984)
for intrinsic solar-like O/H values and above. Castellanos et al. (2002) also
found that the R$_{23}$ -- method yields an overestimated oxygen abundances
in low-excitation H\,{\sc ii} regions. Thus, the high-metallicity part of the 
N/O vs. O/H diagram in previous studies can be highly uncertain. A new way of 
oxygen abundance determination in H\,{\sc ii} regions (called the P -- method) 
was suggested by Pilyugin (2001a). It was demonstrated that the oxygen abundances 
derived with the P -- method are as reliable as the ones derived with the classic 
T$_{e}$ -- method (Pilyugin 2001a,b). Thus, we will be able to construct a 
more realistic  high-metallicity part of the N/O vs. O/H diagram in the 
present study. 

Traditionally, the problem of the nitrogen production is discussed in the 
framework of primary versus secondary processes. 
Since an empirical approach is adopted 
in the present study, the problem of the nitrogen production will be discussed 
in the framework of metallicity-independent versus metallicity-dependent 
yields. 
The metallicity dependence of the yield of element {\it x} does not 
necessary 
imply that element {\it x} is a secondary one. Instead it
 can be result of stellar mass loss which can affect both stellar 
evolution and the production of element {\it x}.
 The dependence of stellar mass loss on metallicity 
leads to a metallicity dependence of the yield of element {\it x} even if 
that element is primary. Indeed, Maeder (1992) predicts a metallicity 
dependence of the yield of the primary element oxygen as the 
result of a metallicity-dependent mass loss process.

The determination of oxygen abundances and nitrogen to oxygen abundance ratios 
in high-metallicity  H\,{\sc ii} regions of galaxies and the construction of an 
accurate N/O -- O/H diagram for H\,{\sc ii} regions in irregular and spiral 
galaxies is  described in Section 2. The nitrogen production sites derived 
from the analysis of the N/O -- O/H diagram and observational data of other 
kinds are reported in Section 3. A discussion is presented in Section 4. 
Section 5 is a brief summary.

\section{The N/O -- O/H diagram}

\subsection{Low-metallicity H\,{\sc ii} regions in irregular galaxies}

Generally, the precision of the oxygen abundance determination in 
oxygen-poor H\,{\sc ii} regions with bright emission lines in irregulars is 
higher than in oxygen-rich H\,{\sc ii} regions in spirals. 
However, many irregular galaxies have no bright H\,{\sc ii} regions with 
readily measured emission lines. As was noted by Hidalgo-G\'amez \& Olofsson 
(1998) the uncertainties in the intensity of the $[OIII] \lambda 4363$ 
line in spectra of H\,{\sc ii} regions in dwarf irregular galaxies reported in 
the literature fluctuate between 11$\%$ and 120$\%$. 
Then the precision of the oxygen abundance determination in oxygen-poor 
irregulars without H\,{\sc ii} regions with bright emission lines is
rather low (Pilyugin 2001c). Therefore we did not collect all the available 
data but instead have considered only two data sets.

\begin{figure}
\resizebox{\hsize}{!}{\includegraphics[angle=0]{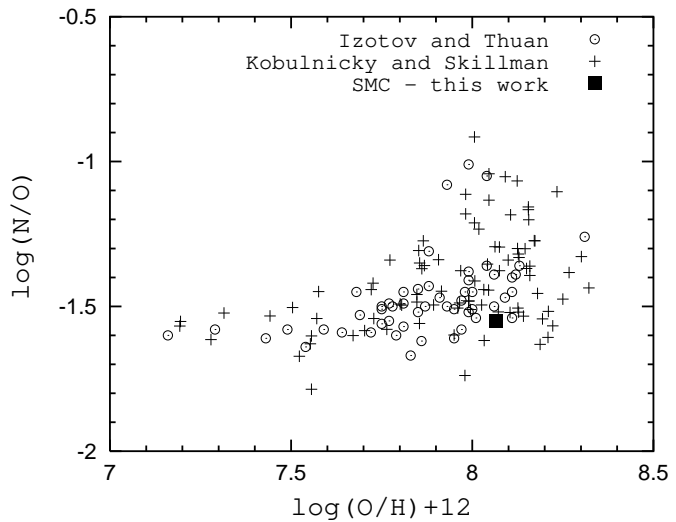}}
\caption{
The N/O versus O/H diagram for H\,{\sc ii} regions in irregular 
galaxies. 
The open circles are data from Izotov \& Thuan (1999), the pluses  are
data from Kobulnicky \& Skillman (1996). The filled square denotes the Small 
Magellanic Cloud (this work). 
}
\label{figure:noz-irr}                                     
   \end{figure}

Kobulnicky \& Skillman (1996) have selected from the literature a large sample 
(72 measurements in 60 galaxies) of spectroscopic measurements in metal-poor 
galaxies (12 + log(O/H) $<$ 
$\sim$ 8.4) and have recomputed the O and N/O abundances in a self-consistent 
manner. Their data are shown  in Fig.\ref{figure:noz-irr} with pluses. 
Izotov \& Thuan (1999) have presented O and N abundances derived from  
high-quality spectroscopic observations of 54 supergiant H\,{\sc ii} regions 
in 50 low-metallicity blue compact galaxies with oxygen abundances 12 + log(O/H) 
between 7.1 and 8.3. These data are presented in Fig.\ref{figure:noz-irr} with 
open circles.

The observed N/O -- O/H diagram in the  
low-metallicity range
has appreciably changed over the last decade. The early 
N/O -- O/H diagram (see Fig.7 in Garnett (1990), Fig.1 in Pilyugin (1992)) 
consisted of a 
cloud of points with a more or less constant upper envelope of N/O, 
and with the spread 
in N/O at a given O/H sligtly decreasing with increasing O/H. 
In contrast, the scatter in N/O at a given O/H in the present  
N/O -- O/H diagram is extremelly small at low O abundances. The scatter 
becomes large only when the metallicity exceeds a certain value 
(12+log(O/H) around 7.6), and up to O abundances around 12+log(O/H) = 8.3 
the lower envelope of N/O is equal to the N/O value at low O abundances, 
Fig.\ref{figure:noz-irr}, (see also Fig.2 in Thuan, 
Izotov, \& Lipovetsky 1995, Fig.15 in Kobulnicky \& Skillman 1996,
Fig.2 in Izotov \& Thuan 1999). 

\subsection{H\,{\sc ii} regions in the Small Magellanic Cloud}

The Small Magellanic Cloud is a well observed dwarf irregular galaxy. Examination of this galaxy can clarify the chemical enrichment history in 
irregulars. The H\,{\sc ii} regions in the Small Magellanic Cloud were observed 
by a number of investigators. Their oxygen and nitrogen abundances were recomputed here in the same way with the  
classic T$_{e}$ -- method, using published spectra. 
The electron temperature within the [OIII] zone is derived from equation 
(Pagel et al. 1992) 
\begin{eqnarray}
t_3 = 1.432/[\log R - 0.85 + 0.03 \log t_3 +  \nonumber  \\ 
\log (1 + 0.0433 x t_{3}^{0.06})] ,
\label{equation:te3}
\end{eqnarray}
where $R$ = $I_{[OIII] \lambda 4959+ \lambda 5007} /I_{[OIII] \lambda 4363}$. 
Using the obtained value of $t_3$, the oxygen and nitrogen abundances 
were derived from the following expressions (Pagel et al. 1992)

\begin{equation}
\frac{N}{O} = \frac{N^+}{O^+}  ,
\label{equation:nono}
\end{equation}
\begin{eqnarray}
\log (N^+/O^+) = \log \frac{I_{[NII] \lambda 6548 + \lambda 6584}}
{I_{[OII] \lambda 3726 + \lambda 3729}} + 0.307 - \frac{0.726}{t_2}  \nonumber  \\ 
 - 0.02 \log t_2  , 
\label{equation:no}
\end{eqnarray}
\begin{equation}
\frac{O}{H} = \frac{O^+}{H^+} + \frac{O^{++}}{H^+}                ,
\label{equation:otot}
\end{equation}
\begin{eqnarray}
12+ \log (O^{++}/H^+) = \log \frac{I_{[OIII] \lambda 4959 + \lambda 5007}}
{I_{H_{\beta}}} + 6.174 +  \nonumber  \\ 
\frac{1.251}{t_3}  - 0.55 \log t_3 ,
\label{equation:oplus2}
\end{eqnarray}
\begin{eqnarray}
12+ \log (O^{+}/H^+) = \log \frac{I_{[OII] \lambda 3726 + \lambda 3729}}
{I_{H_{\beta}}} + 5.890 +  \nonumber  \\ 
\frac{1.676}{t_2}  - 0.40 \log t_2 + \log (1+1.35x)  ,
\label{equation:oplus}
\end{eqnarray}
\begin{equation}
x= 10^{-4} n_e t_2^{-1/2}, 
\label{equation:x}
\end{equation}
where $n_e$ is the electron density in cm$^{-3}$,  $t_2$ = $t_{[NII]}$ 
is the electron temperature in units of 10$^4$K. We adopt $t_2$ = 
$t_{[NII]}$  = $t_{[OII]}$. 
The value of $n_e$ is adopted to be equal to 100 cm$^{-3}$ for all  
H\,{\sc ii} regions. 
The $t_2$ -- $t_3$ relation was taken from Garnett (1992):
\begin{equation}
t_2 =  0.7 \, t_3 + 0.3 .
\label{equation:t2t3}
\end{equation}

The results are presented in Table \ref{table:smc}. 
The name of the H\,{\sc ii} region is listed in column 1. The oxygen 
abundance is reported in column 2. The nitrogen to oxygen abundance ratio
is given in column 3. The reference is listed in column 4. Examination of 
Table \ref{table:smc} shows that the oxygen and nitrogen abundances in the 
interstellar medium of the Small Magellanic Cloud are well defined; 
oxygen and nitrogen abundances derived with spectroscopic data from 
different studies are in good agreement. We note that the nitrogen 
to oxygen ratio in the H\,{\sc ii} regions of the Small Magellanic Cloud 
is very close to that in the H\,{\sc ii} regions of the most oxygen-poor blue 
compact galaxies, Fig.\ref{figure:noz-irr}. 

\begin{table}
\caption[]{\label{table:smc}
The oxygen and nitrogen abundances in H\,{\sc ii} regions of the Small 
Magellanic Cloud recomputed here with the classic T$_{e}$ -- method 
using published spectra. 
The name of the H\,{\sc ii} region is listed in column 1. The oxygen 
abundance is reported in column 2. The nitrogen to oxygen abundance ratio
is given in column 3. The reference is listed in column 4. 
}
\begin{center}
\begin{tabular}{llll} \hline \hline
                   &            &            &                  \\  
H\,{\sc ii} region & 12+log(O/H)  & log(N/O)   & reference        \\  
                   &            &            &                  \\   \hline
N66a               & 8.081      &  -1.525    & Duf75            \\   
N83a               & 8.122      &  -1.582    & Duf75            \\   
NGC346 I           & 8.034      &  -1.520    & PTP76            \\   
NGC346 II          & 8.068      &  -1.633    & PTP76            \\   
NGC356 I           & 7.980      &  -1.678    & PTP76            \\   
NGC456 I           & 8.087      &  -1.502    & PTP76            \\   
N12a               & 8.156      &  -1.509    & DH77             \\   
N12b               & 8.018      &  -1.590    & DH77             \\   
N22                & 8.206      &  -1.665    & DH77             \\   
N25                & 8.037      &  -1.551    & DH77             \\   
N66NW              & 8.098      &  -1.603    & DH77             \\   
N66SE              & 8.114      &  -1.465    & DH77             \\   
N76E               & 7.986      &  -1.321    & DH77             \\   
N76SW              & 7.902      &  -1.418    & DH77             \\   
N81                & 8.120      &  -1.513    & DH77             \\   
N83                & 8.105      &  -1.597    & DH77             \\   
N88                & 7.971      &  -1.591    & DH77             \\   
N90                & 8.094      &  -1.546    & DH77             \\   
N13                & 8.117      &  -1.547    & Pag78            \\   
NGC346 A           & 8.056      &  -1.598    & PPR00            \\   
NGC346 B           & 8.033      &  -1.593    & PPR00            \\   
{\sc mean}         & 8.066$\pm$0.068 &  -1.550$\pm$0.080 &      \\     \hline  \hline 
\end{tabular}
\end{center}

\vspace{0.05cm}

{\it List of references}:

DH77   --  Dufour \& Harlow 1977;
Duf75  --  Dufour 1975;
Pag78  --  Pagel et al. 1978; 
PPR00  --  Peimbert, Peimbert, Ruiz 2000;
PTP76  --  Peimbert \& Torres-Peimbert 1976
\end{table}

\subsection{High-metallicity H\,{\sc ii} regions in spiral galaxies}

The N/O vs. O/H diagram for H\,{\sc ii} regions in spiral galaxies has been 
constructed 
in a number of studies (Matteucci 1986; Vila-Costas \& Edmunds 1993;  
van Zee et al. 1998a; Henry et al. 2000; among others). In those studies, the 
oxygen abundances in high-metallicity H\,{\sc ii} regions of spiral galaxies 
are derived using an empirical R$_{23}$ -- calibration, 
inducing a large uncertainty in the N/O vs. 
O/H diagram as mentioned in the Introduction. 
Here we attempt to construct a more precise  
N/O vs. O/H diagram for H\,{\sc ii} regions in spiral galaxies. The oxygen 
abundances and N/O abundance ratios in high-metallicity H\,{\sc ii} regions, 
where $\lambda 4363$ is not available,  are determined with the following 
algorithm. The oxygen abundance is determined from the expression (Pilyugin 
2001a)
\begin{equation}
12+\log(O/H)_{P} = \frac{R_{23} + 54.2  + 59.45 P + 7.31 P^{2}}
                       {6.07+6.71P+0.371P^{2}+0.243R_{23}} .
\label{equation:ohp}
\end{equation}
The following notations have been adopted here:
$R_{2}$ = $I_{[OII] \lambda 3727+ \lambda 3729} /I_{H\beta }$, 
$R_{3}$ = $I_{[OIII] \lambda 4959+ \lambda 5007} /I_{H\beta }$, 
$R_{23}$ =$R_{2}$ + $R_{3}$, P=R$_3$/R$_{23}$.

Using the value of O/H derived from Eq.(\ref{equation:ohp}) and measured line 
intensities, Eqs.(\ref{equation:otot}) -- (\ref{equation:t2t3}) can be solved 
for $t_{2}$.  Then the N/O abundance ratio is derived from the 
Eqs.(\ref{equation:nono}) -- (\ref{equation:no}). 

Alternatively, the value of $t_{2}$ can be also found from the following 
expression for $t_P$ = $t_3$ (Pilyugin 2001a)
\begin{equation}
t_{P} = \frac{R_{23} + 3.09  + 7.05 P + 2.87 P^{2}}
                       {9.90  + 11.86 P + 7.05 P^{2} - 0.583 R_{23}}  
\label{equation:tp}
\end{equation}
and Eq.(\ref{equation:t2t3}). The value of $t_2$ determined from
Eqs.(\ref{equation:t2t3}),(\ref{equation:tp}) is in good agreement 
with the value of $t_2$ derived from
Eqs.(\ref{equation:otot})-(\ref{equation:ohp})
for the majority of the H\,{\sc ii} regions considered. However, for H\,{\sc
ii} regions in which most of the oxygen is in the O$^+$ stage, 
the two values
of $t_2$ are not in agreement. Eqs.(\ref{equation:otot})-(\ref{equation:ohp})
give more realistic values of $t_{2}$ for these H\,{\sc ii} regions.

\begin{figure}
\resizebox{\hsize}{!}{\includegraphics[angle=0]{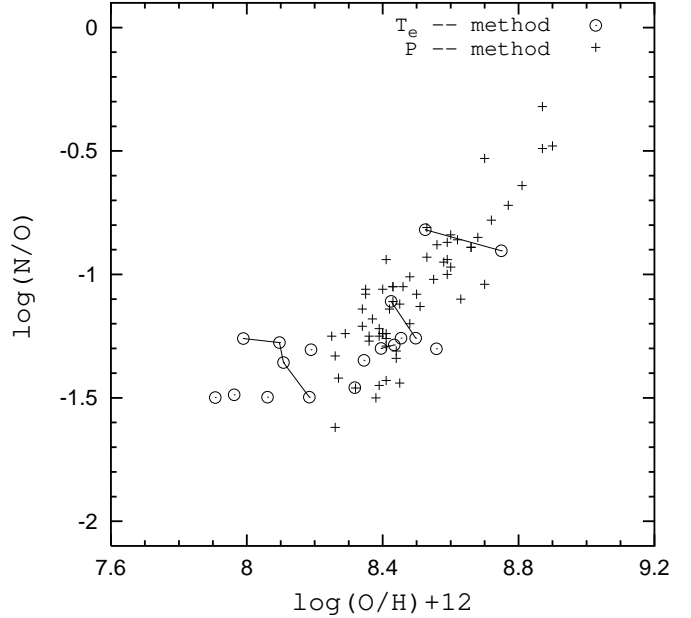}}
\caption{
The (N/O)$_{T_{e}}$ -- (O/H)$_{T_{e}}$ diagram for H\,{\sc ii} regions of M101 
with  measured temperature-sensitive line ratios is shown by open 
circles. Two or 
more independent observations of the same H\,{\sc ii} region
 are connected with a solid line. 
The (N/O)$_{P}$ -- (O/H)$_{P}$ diagram for H\,{\sc ii} regions of 
M101 is shown by the pluses. 
}
\label{figure:noz-m101}                                     
   \end{figure}

The comparison between  the 
(N/O)$_{T_{e}}$ -- (O/H)$_{T_{e}}$ and (N/O)$_{P}$ -- 
(O/H)$_{P}$ diagrams for H\,{\sc ii} regions of the well -- observed spiral 
galaxy M101 allows to test the reliability of the (N/O)$_{P}$ values.
The spectra of the H\,{\sc ii} regions in M101 with  measured temperature-sensitive 
line ratios are given in a number of publications 
(Garnett \& Kennicutt 1994; Garnett et al. 1999; Kinkel \& Rosa 1994;
McCall et al. 1985; Rayo et al. 1982; Shields \& Searle 1978; Smith 1975; 
Torres-Peimbert et al. 1989; van Zee et al. 1998b). 
The (N/O)$_{T_{e}}$ -- (O/H)$_{T_{e}}$ diagram for H\,{\sc ii} regions of 
M101 with  measured temperature-sensitive line ratios is presented in 
Fig.\ref{figure:noz-m101} by open circles. Two or more independent 
observations of the same H\,{\sc ii} region are connected 
by a solid line. 
The (N/O)$_{P}$ -- (O/H)$_{P}$ diagram for the H\,{\sc ii} regions of 
M101 is shown in Fig.\ref{figure:noz-m101} by the pluses. 
The intensities of $I_{[OII] \lambda 3727+ \lambda 3729} /I_{H\beta }$,  
$I_{[OIII] \lambda 4959+ \lambda 5007} /I_{H\beta }$, and 
$I_{[NII] \lambda 6548 + \lambda 6584} /I_{H\beta }$  were taken from 
Smith (1975), Shields \& Searle (1978), Rayo, Peimbert, \& Torres-Peimbert 
(1982), McCall, Rybski, \& Shields (1985), Torres-Peimbert, Peimbert, \& Fierro 
(1989), Garnett \& Kennicutt (1994), Kinkel \& Rosa (1994), Kennicutt \& Garnett 
(1996), van Zee et al. (1998b), Garnett et al. (1999).
Inspection of Fig.\ref{figure:noz-m101} shows that there is good agreement 
between the (N/O)$_{T_{e}}$ -- (O/H)$_{T_{e}}$ and (N/O)$_{P}$ -- 
(O/H)$_{P}$ diagrams, confirming the reliability of the (N/O)$_{P}$ and 
(O/H)$_{P}$ values.

\begin{figure}
\resizebox{\hsize}{!}{\includegraphics[angle=0]{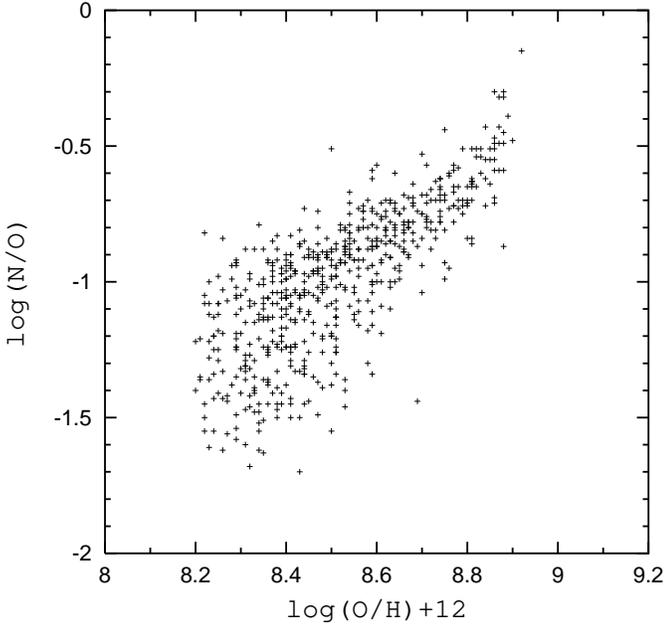}}
\caption{
The N/O -- O/H diagram for H\,{\sc ii} regions in spiral galaxies.
}
\label{figure:noz-spir}                                     
   \end{figure}

The oxygen abundances (O/H)$_{P}$ and nitrogen-to-oxygen abundances ratios 
(N/O)$_{P}$ have been derived for H\,{\sc ii} regions in the 
following galaxies:
 NGC224, NGC253, NGC300, NGC598,  NGC628,  NGC753,  NGC925,  NGC1058, 
 NGC1232,  NGC1365,  NGC1637,  NGC2403,  NGC2805, NGC2841,  NGC2903, 
 NGC2997,  NGC3031,  NGC3184,  NGC3351,  NGC4254,  NGC4258,  NGC4303, 
 NGC4321,  NGC4395,  NGC4501, NGC4751,  NGC4651,  NGC4654,  NGC4689, 
 NGC4713,  NGC4736,  NGC5055,  NGC5194,  NGC5236,  NGC5457,  NGC6384, 
 NGC6946,  NGC7331,  NGC7793,  and IC342.
We use Eqs.(\ref{equation:nono}) - (\ref{equation:ohp}) and 
more than six hundred published spectra from        
        Alloin et al. (1981), 
        Blair et al. (1982), 
        Bresolin et al. (1999), 
        Dennefeld \& Kunth (1981), 
        Diaz et al. (1991), 
        Dufour et al. (1980), 
        Edmunds \& Pagel (1984), 
        Ferguson et al. (1998), 
        Fierro et al. (1986), 
        Henry et al. (1992,1994, 1996), 
        Garnett et al. (1997), 
        Garnett et al. (1999), 
        Garnett \& Shields (1987), 
        Kennicutt \& Garnett (1996), 
        Kinkel \& Rosa (1994), 
        Kwitter \& Aller (1981), 
        McCall et al. (1985), 
        Pagel et al. (1979), 
        Rayo et al. (1982), 
        Roy \&  Walsh (1997), 
        Searle (1971), 
        Shields et al. (1991), 
        Skillman et al. (1996), 
        Smith (1975), 
        Staufer \& Bothum (1984), 
        Torres-Peimbert et al. (1989), 
        van Zee et al. (1998b), 
        Vilchez et al. (1988), 
        and Webster \& Smith (1983). 
The N/O -- O/H diagram for H\,{\sc ii} regions in spiral galaxies is shown  
in Fig.\ref{figure:noz-spir}. The detailed discussion of the radial nitrogen 
abundance distribution across the disks of spiral galaxies will be given 
elsewhere.

\begin{figure}
\resizebox{\hsize}{!}{\includegraphics[angle=0]{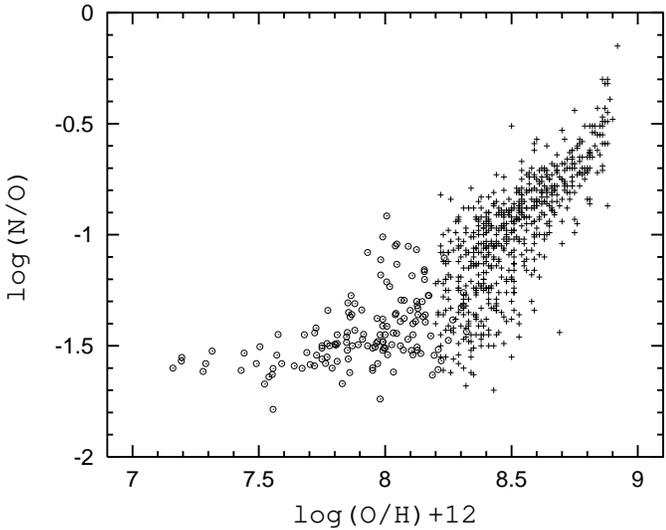}}
\caption{
The N/O -- O/H for H\,{\sc ii} regions in spiral (pluses) and irregular 
(circles) galaxies. 
}
\label{figure:noz-tot}                                     
   \end{figure}

A prominent feature of Fig.\ref{figure:noz-spir} is the lack of H\,{\sc ii} 
regions with 12+log(O/H) lower 8.2. This lack does not mean that there are no 
low-metallicity H\,{\sc ii} regions in the outer parts of spiral galaxies. 
Rather, it is caused by the following reason:   
the relationship between oxygen abundance and strong line intensities is double-valued with two distinctive parts usually known as 
the lower and upper branches
of the R$_{23}$ -- O/H diagram. Thus, one has to know in advance on which of
the two branches the H\,{\sc ii} region lies. 
The above
expression for the oxygen abundance determination in H\,{\sc ii} regions,
Eq.(\ref{equation:ohp}), is only 
valid for H\,{\sc ii} regions which belong to the 
upper branch, with 12+log(O/H) higher
than around 8.2.
It has been known for a long time (Searle 1971, Smith 1975) that disks of spiral
galaxies can show radial oxygen abundance gradients, in the sense that
the oxygen abundance is higher in the central part of the disk and decreases
with galactocentric distance. 
Thus, we start from the H\,{\sc ii} regions in the central part of disk and 
move outward until the radius R$^*$ where the oxygen 
abundance decreases to around 12+log(O/H) = 8.2. An unjustified use 
of Eq.(\ref{equation:ohp}) in the determination of the 
oxygen abundance in low-metallicity H\,{\sc ii} regions beyond R$^*$ 
would result in overestimated oxygen abundances, and would cause a  
false bend in the slope of abundance gradients to appear. 
Therefore, H\,{\sc ii} regions with galactocentric
distances larger than R$^*$, those with 12+log(O/H) less than 8.2 
were rejected. 

The general N/O -- O/H diagram for H\,{\sc ii} regions in spiral and irregular 
galaxies is shown in Fig.\ref{figure:noz-tot}.

\section{The nitrogen production sites traced by the observational data}

\subsection{The N/O -- O/H diagram}

The observed trend of nitrogen abundance with metallicity as 
measured by oxygen 
abundance has been widely used as a clue to identify the sites for nitrogen 
production (Edmunds \& Pagel 1978; Garnett 1990; Pilyugin 1992, 1993; 
Marconi, Matteucci, \& Tosi 1994; Thuan, Izotov, \& Lipovetsky 1995; Izotov \& 
Thuan 1999; Henry, Edmunds, \& K\"oppen 2000; among others). Conclusions 
about the nitrogen production sites are reached from the interpretation of 
 three prominent 
features in the observed N/O -- O/H diagram : 
{\it i)} the remarkable constancy of N/O in the lowest metallicity (12+log(O/H) $<$ 
7.6) galaxies, and the fact that, up to O abundances around 12+log(O/H) = 8.3 
the lower envelope of N/O is equal to the N/O value at low O abundances; 
{\it ii)} beginning at roughly 12+log(O/H) = 7.6, there is a large scatter in N/O at 
a given O/H value;
{\it iii)} beginning at roughly 12+log(O/H) = 8.3, the lower envelope of N/O 
increases with O/H. 

In the early 1990s, several studies were devoted to the explanation of the 
N/O -- O/H diagram (Garnett 1990; Pilyugin 1992, 1993; Vila-Costas \& Edmunds
1992, Marconi, Matteucci, \& Tosi 1994). The problem of the scatter in N/O at a 
given O/H value has attracted particular attention. Two mechanisms have 
been suggested to explain the scatter. The first 
mechanism is time-delayed production of nitrogen. The second mechanism is 
selective heavy-element loss through enriched galactic winds. 
As mentioned above, the observed present-day N/O -- O/H diagram at low 
metallicities differs significantly from the earlier one: while the earlier  
diagram suggests that the scatter in N/O grows as O/H 
decreases, the present one shows exactly 
the reverse. Therefore the interpretation suggested in the early 1990s has 
been revised in later studies. 

\begin{figure*}
\resizebox{\hsize}{!}{\includegraphics[angle=0]{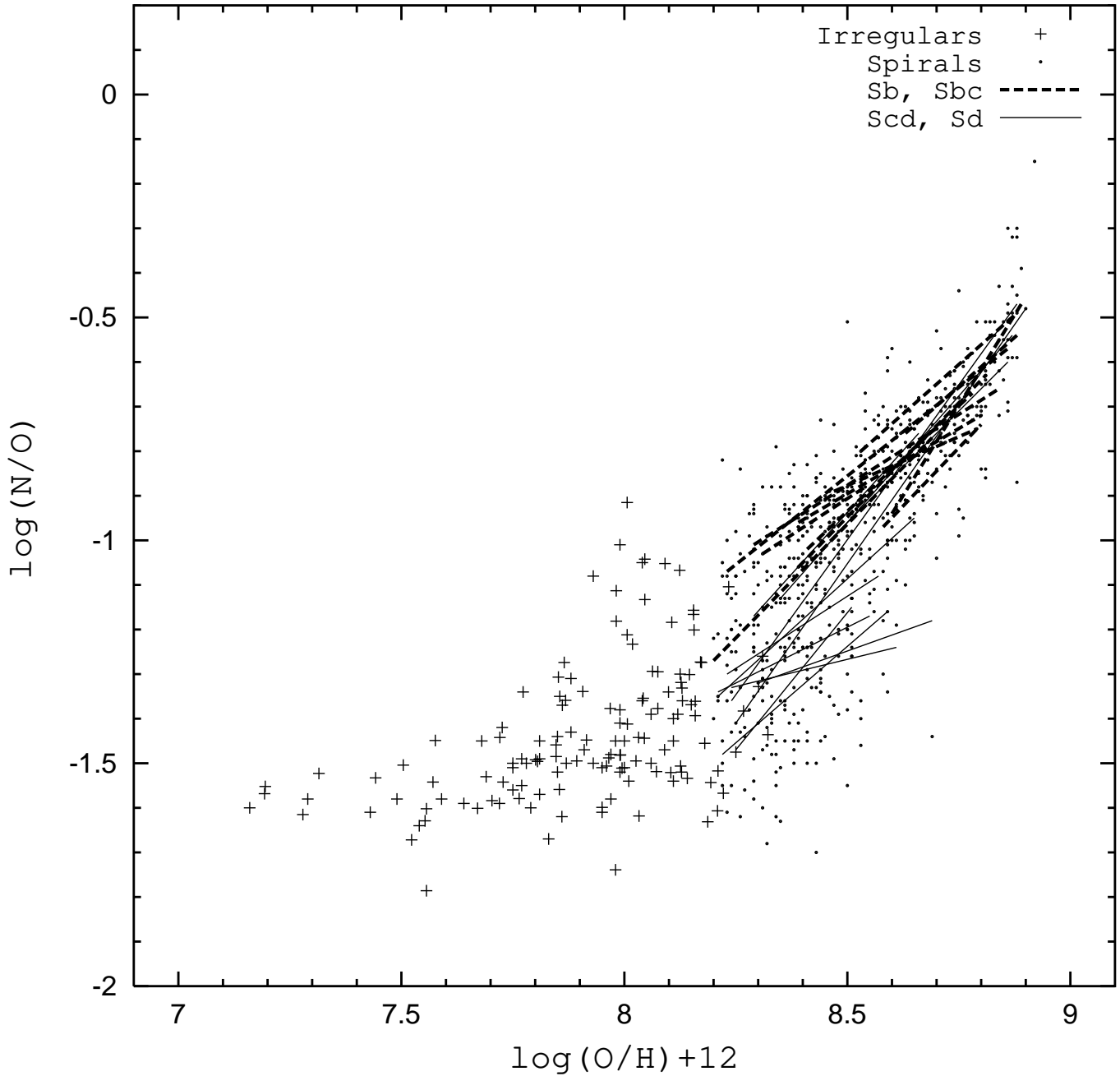}}
\caption{
The N/O -- O/H diagram. The pluses are H\,{\sc ii} regions in irregular galaxies.
The points are H\,{\sc ii} regions in spiral galaxies. The thick dashed lines are 
the N/O -- O/H trends in Sb and Sbc galaxies (T-type = 3,4). The thin solid 
lines are the N/O -- O/H trends in Scd and Sd galaxies (T-type = 6,7). 
}
\label{figure:noz-type}                                     
   \end{figure*}
 
Thuan, Izotov \& Lipovetsky (1995) and Izotov \& Thuan (1999) have concluded 
that the small dispersion of N/O at low metallicities is strong evidence against 
any time-delayed production of nitrogen in the lowest metallicity blue compact 
galaxies and the constancy of N/O can be explained only by nitrogen production 
in short-lived massive stars, i.e. oxygen and nitrogen must be made in the 
same massive stars. Nitrogen production by massive stars was also suggested 
by Matteucci (1986). 
Thuan, Izotov, \& Lipovetsky (1995) and Izotov \& Thuan 
(1999) have concluded that the large dispersion of N/O in galaxies with 
12+log(O/H) $>$ 7.6 can be explained by the addition of nitrogen production 
in intermediate-mass stars; by the time intermediate-mass stars have evolved 
and released their nucleosynthetic products (100 -- 500 Myr), all galaxies have 
become enriched to 7.6 $<$ 12 + log(O/H) $<$ 8.2 -- the delayed release of 
nitrogen greatly increase the scatter in N/O. Close examination of the 
N/O -- O/H diagram shows that this scenario for the 
history of galaxy enrichment in 
nitrogen meets difficulties in galaxies 
with 12+log(O/H) $>$ 7.6. Indeed, the N/O value in the Small Magellanic Cloud 
is close to that in the lowest metallicity blue compact galaxies, 
Fig.\ref{figure:noz-irr}, (see below). According to the above scenario, 
only massive stars are responsible for the N/O value in the Small 
Magellanic Cloud. Then one has to conclude that most of the  
stars in the Small Magellanic Cloud have formed recently so that the 
nitrogen-producing stars have not yet returned their nucleosynthesis products 
to the interstellar medium because they have not had enough time to evolve.   
In that scenario, most of the stars in the Small Magellanic 
Cloud formed in last 100 -- 500 Myr, which is not the case. 

Henry, Edmunds, \& K\"oppen (2000) have concluded that the constancy of 
the N/O in metal-poor galaxies is entirely and naturally explained if they 
are characterized by historically low star formation rates, with nearly 
all of the nitrogen being produced by intermediate mass stars between 4 
and 8 M$_{\odot}$, with a moderate time delay (a characteristic lag time of 
roughly 250 Myr following their formation). 
Those authors have found that this moderate time delay 
in nitrogen release by intermediate-mass stars 
does not appear to be an important factor 
in the evolution of this element. They concluded that the scatter is caused 
by intermittent increases in nitrogen caused by local contamination by 
Wolf-Rayet stars or luminous blue variables. Given that the 
scatter is large (more than 0.3 dex at 12+log(O/H) = 7.9), 
this conclusion appears to be in conflict 
with their other conclusion that nearly all of the nitrogen is produced by 
intermediate-mass stars between 4 and 8 M$_{\odot}$. 

\begin{figure}
\resizebox{\hsize}{!}{\includegraphics[angle=0]{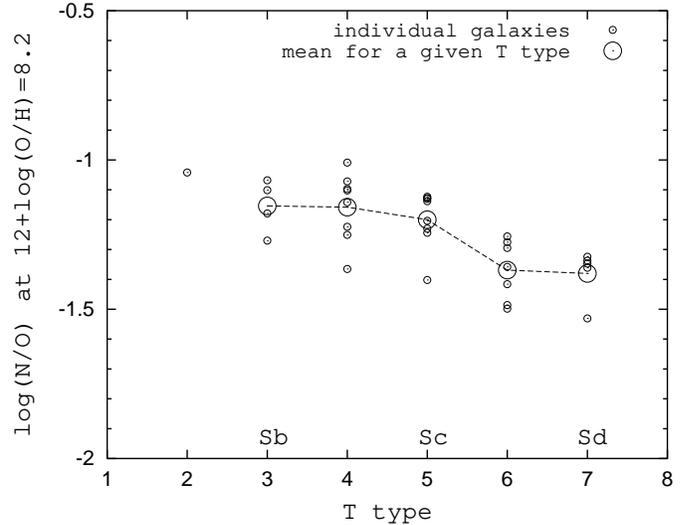}}
\caption{The small circles are the log(N/O) values at 12+log(O/H)=8.2 in 
individual galaxies as a function of the T-type. The large circles are  
the log(N/O) values at 12+log(O/H)=8.2 averaged over all  
galaxies of the same T-type.
}
\label{figure:no-oh82}                                     
   \end{figure}

Thus, the existing explanations of the large scatter in N/O at a given O/H 
do not appear to be satisfactory. We have found that the low-metallicity 
H\,{\sc ii} regions in the outer parts of spiral galaxies are more suitable 
objects for establishing
the origin of the scatter than H\,{\sc ii} regions in irregular 
galaxies. Our finding is the following. 
The positions of spiral galaxies of different morphological types in the 
N/O -- O/H diagram are shown in Fig.\ref{figure:noz-type} by different symbols. 
The dashed lines correspond to Sb and Sbc galaxies (T-type = 3,4), and 
the solid 
lines to Scd and Sd galaxies (T-type = 6,7). Fig.\ref{figure:noz-type} 
shows clearly that the N/O values in H\,{\sc ii} regions of galaxies of early 
morphological types are systematically higher than the N/O values in H\,{\sc ii} regions 
with the same O/H value in galaxies of late morphological types. It can be 
also seen in Fig.\ref{figure:no-oh82} where the N/O ratio at 12 + log(O/H) = 
8.2 is shown as a function of numerical morphological T-type.
It is interesting to note that although Vila-Costas \& Edmunds (1992) have 
concluded that there is 
no obvious difference between morphological types, 
a hint that the N/O ratios in the H\,{\sc ii} regions of the late-type spiral 
galaxies are lower than the ones in the H\,{\sc ii} regions with the 
same oxygen abundance 
in early-type spiral galaxies can be seen in their Fig.5. The much more 
distinct correlation of the N/O ratio with galaxy morphological type in the  
present study is due to the fact that the positions of H\,{\sc ii} regions in the 
N/O -- O/H diagram derived here with the P -- method are more accurate than 
those derived by Vila-Costas \& Edmunds (1992). 

\begin{figure}
\resizebox{\hsize}{!}{\includegraphics[angle=0]{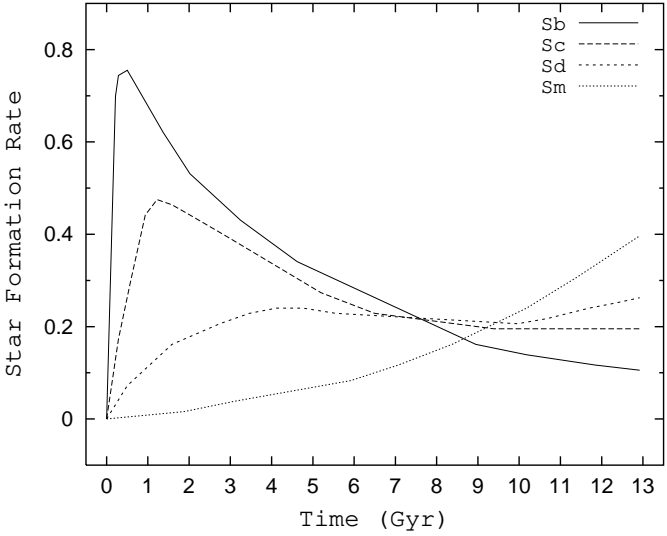}}
\caption{
The schematic star formation history of galaxies of different morphological 
types according to Sandage (1986). The star formation rate is in arbitrary 
units.
}
\label{figure:sandage}                                     
   \end{figure}

It is well-known that galaxies of different morphological types have 
different star formation histories. The schematic star formation histories of 
galaxies of different morphological types according to 
Sandage (1986) are presented 
in Fig.\ref{figure:sandage}. The star formation rate is in arbitrary units.
Fig.\ref{figure:sandage} shows that the spiral galaxies with early 
morphological types have a significantly larger fraction of old stars (with 
ages 8 - 13 Gyr) than the galaxies of late morphological types. This suggests 
that the high N/O values in the galaxies of early morphological types can be 
caused by the contribution of low-mass stars (with masses around a 
solar mass) 
to the nitrogen production, while in the galaxies of late morphological types
these stars have not yet returned their nucleosynthesis products 
to the interstellar medium because they have not had enough time to evolve.   
The long time delay for nitrogen production appears to be an important factor in 
the evolution of this element. The scatter in N/O values at a given O/H can 
be naturally explained by differences in star formation histories in galaxies. 
This conclusion is not new. More than twenty 
years ago, Edmunds \& Pagel (1978) have suggested that observations of the 
N/O abundance ratio in external galaxies can be understood if nitrogen is 
manufactured principally in stars of 1 -- 2.5 M$_{\odot}$. The N/O ratio of a 
galaxy then becomes an indicator of the time that has elapsed since the bulk 
of star formation occurred, or in other words of 
the nominal "age" of the galaxy.

The explanation that the increase of the lower envelope of N/O with O/H, 
beginning roughly at 12 + log(O/H) = 8.3, is controlled by the 
metallicity-dependent nitrogen production in intermediate-mass stars is 
commonly accepted now.

Thus, the present and previous studies of N/O -- O/H diagram lead to the 
following conclusions about the sites of nitrogen production: 
(1) at low metallicities the metallicity-independent nitrogen production takes 
place in short-lived massive stars with no time delay between the release of 
nitrogen and oxygen or/and in intermediate-mass stars with moderate time delay, 
with a characteristic time delay of a few hundred Myr between the release of 
nitrogen and oxygen; 
(2) at metallicities higher than roughly 12 + log(O/H) = 8.3, the 
metallicity-dependent nitrogen production by intermediate-mass stars dominates; 
(3)  the contribution of low-mass stars to the nitrogen production with long 
time delay, with characteristic time delay of several Gyr, can be significant. 

Observational data of other kinds can be used to verify and  
refine conclusions reached by us and other authors from consideration of the 
N/O -- O/H diagram. 

\subsection{The self-enriched H\,{\sc ii} regions}

Generally, stars of 
different masses from all the previous star formation events can make 
contributions to the N abundances measured in the interstellar medium. 
Fortunately, there are cases where the contribution to nitrogen production 
from massive stars from a current 
star formation event can be established. 
Kunth \& Sargent (1986) have 
suggested that H\,{\sc ii} regions may "pollute" themselves
on short (10$^6$ yr) timescales with nucleosynthetic products from the current 
burst of star formation. This suggestion is not beyond question. There may 
be some time lag ($\sim$ 10$^7$ yrs) between supernova explosions and 
the appearance of the freshly produced oxygen in the 
warm phase of the interstellar 
gas (Tenorio-Tagle 1996; Kobulnicky \& Skillman 1996, 1997).
There are however examples where the enrichment in heavy elements by 
massive stars on short timescales is exhibited.
Those examples can tell us something about the nitrogen production by massive 
stars. 

\begin{figure}
\resizebox{1.00\hsize}{!}{\includegraphics[angle=0]{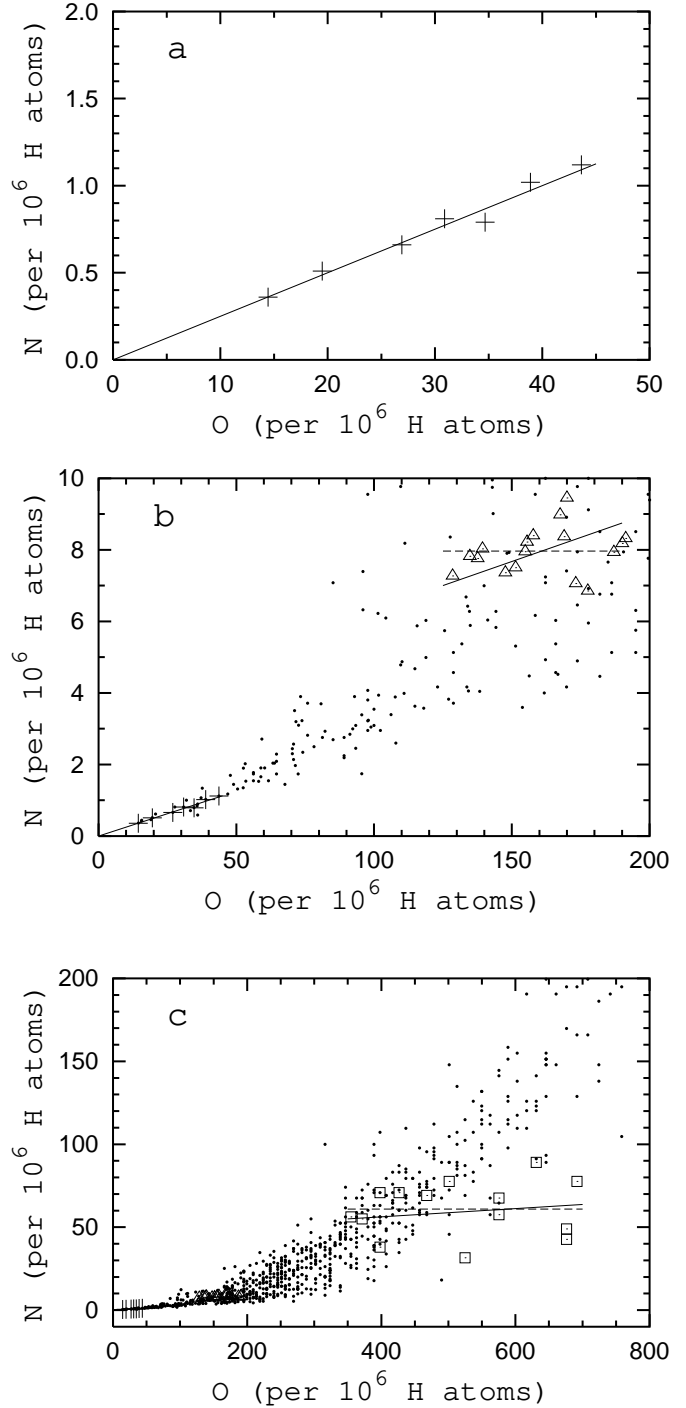}}
\caption{
Examples in which the enrichment in heavy elements by massive stars is expected.
{\bf a)}. Pluses are H\,{\sc ii} regions in extremely metal-poor blue compact dwarf
galaxies from Izotov \& Thuan (1999). The solid line is the best fit to them. 
{\bf b)}. 
Triangles are distinct spatial locations of multiple starburst 
knots within the galaxy NGC 4214 (Kobulnicky \& Skillman 1996). 
Points are the same objects as in Fig.\ref{figure:noz-tot}.
The solid line is the track corresponding to the ratio of nitrogen 
to oxygen enrichment yields $\Delta$N = $\frac{1}{40} \Delta$O.
The dashed line corresponds to $\Delta$N = 0.
{\bf c)}. Squares are B stars from the Orion association (Cunha \& Lambert 1994). 
Points are the same objects as in Fig.\ref{figure:noz-irr}.
The solid line is the track corresponding to a ratio of nitrogen 
to oxygen enrichment yields $\Delta$N = $\frac{1}{40} \Delta$O.
The dashed line corresponds to $\Delta$N = 0.
}
\label{figure:polution}                                     
   \end{figure}

Let us test the conclusion of Izotov \& Thuan (1999) that the nitrogen in 
low-metallicity (with 12+log(O/H) $<$ 7.6) blue compact galaxies is produced 
by massive stars only. The best fit to Izotov \& Thuan's data results in 
$\Delta$N/$\Delta$O $\sim$ 1/40 (solid line in Fig.\ref{figure:polution}a).

Kobulnicky \& Skillman (1996) carried out an investigation of the multiple 
starburst knots in the nucleus of the low-metallicity irregular galaxy NGC4214. 
From optical spectroscopy at distinct spatial locations they reveal 
large scale (about 200 pc) variations in the oxygen abundance. The southernmost, 
and apparently youngest emission line region exhibits a higher oxygen abundance 
than the rest of the galaxy. The N/O ratio is lower in this region. 
Kobulnicky \& Skillman suggest that recent oxygen polution from supernovae 
has occurred in this very young population of massive stars. Their data 
(taken from their Fig.8) are shown in Fig.\ref{figure:polution}b by triangles.
The variation of oxygen and nitrogen abundances computed with 
$\Delta$N = $\frac{1}{40} \Delta$O is shown by the solid line.
Examination of Fig.\ref{figure:polution}b shows that 
Kobulnicky and Skillman's 
data agree satisfactory with the prediction of the model with 
$\Delta$N = $\frac{1}{40} \Delta$O (solid line), however these data 
agree equally satisfactory with the prediction of the model without nitrogen 
production by massive stars, $\Delta$N = 0, (dashed line). 
 
Cunha \& Lambert (1992, 1994) have measured the N, C, O, Si, and Fe abundances 
of main-sequence B stars in the Orion star formation region. There is a time 
lag of about 11 Myr between the formation of the oldest and the youngest 
subgroups of stars in the Orion association. The stars in the youngest subgroup 
are enriched in heavy elements relative to the stars belonging to the oldest 
subgroup, i.e. the later subgeneration of stars have formed out of material 
whose chemical composition has been altered by the nucleosynthesis products
 of massive stars from previous subgenerations. 
Again, the track with $\Delta$N/$\Delta$O $\sim$ 1/40 (solid line) and 
the track with $\Delta$N = 0 (dashed line) cannot be chosen on the basis 
of observational data (Fig.\ref{figure:polution}c). 

Thus, the consideration of the H\,{\sc ii} regions taken to be self-enriched
does not result in a firm argument pro or contra nitrogen production by 
massive stars.

\subsection{The planetary nebulae}

It is usually accepted that the oxygen abundances measured in planetary nebulae 
reflect the initial abundance of the progenitor stars while the nitrogen 
abundances in planetary nebulae with progenitors from some range of stellar mass 
are modified by the progenitor's evolution. Then the oxygen abundance in the 
planetary nebula can serve as some kind of "age" indicator of the progenitor if 
the chemical evolution of the galaxy (the history of galaxy enrichment in oxygen) 
is established. It should be stressed that we do not imply to derive the age 
of the progenitor of individual planetary nebula based on its oxygen abundance. 
Indeed, the temporal evolution of the oxygen abundance in our Galaxy is nonsmooth
(Pilyugin \& Edmunds 1996a,b), therefore the oxygen abundance in the individual 
planetary nebula is not a reliable "age" indicator of its progenitor. 
We only mean that the assumption, that the oxygen-rich planetary nebulae 
(with 12+logO/H $>$ 7.7 in the case of the Small Magellanic Cloud) were formed 
from younger (in average) and therefore more massive stars than the oxygen-poor 
planetary 
nebulae (with 12+logO/H $<$ 7.5 in the case of the Small Magellanic Cloud), is 
justified. Then the examination of nitrogen abundances of planetary nebulae 
with different oxygen abundances allows to make conclusion about the contribution 
of the intermediate- and low-mass stars to the enrichment of the interstellar 
medium in nitrogen. 

The determinations of the chemical composition in a number of planetary nebulae 
of the Small Magellanic Cloud are available now (Richer 1993, Leisy \& Dennefeld 
1996, Costa et al. 2000). This data allows 
to make conclusion about the contribution of the intermediate- and low-mass 
stars to the enrichment of the interstellar medium in nitrogen. Richer (1993) 
reported the oxygen and nitrogen abundances for twenty-five planetary nebulae in  
the Small Magellanic Cloud, Leisy \& Dennefeld (1996)
for fifteen planetary nebulae, and Costa et al. (2000) 
for twenty-one planetary nebulae 
 
The list of Richer (1993) has ten planetary nebulae in common with the list 
of Leisy \& Dennefeld (1996) and fifteen planetary nebulae in common with 
the list of Costa et al. (2000). 
Comparison of the oxygen abundances in common planetary nebulae shows that 
the ones derived by Costa et al. (2000) and Leisy \& Dennefeld 
(1996) are shifted towards higher oxygen abundances when compared to those 
derived by Richer (1993). In order to have a homogeneous set of 
abundances, the abundances of both oxygen and nitrogen of Costa et al. and of 
Leisy \& Dennefeld were reduced to the system of Richer. We chose  
Richer's system because his planetary nebulae oxygen abundances 
are in better agreement with those in 
H\,{\sc ii} regions. 

\begin{figure}
\resizebox{\hsize}{!}{\includegraphics[angle=0]{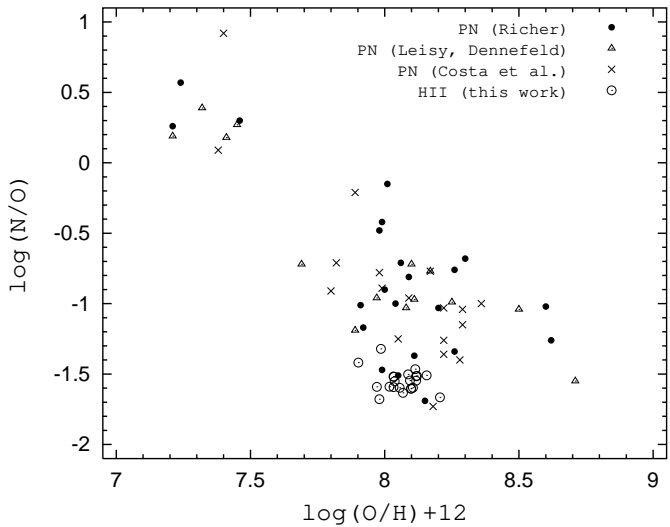}}
\caption{
The O/H - N/O diagram for H\,{\sc ii} regions and planetary nebulae in the Small 
Magellanic Cloud. The points are original data for planetary nebulae from 
Richer (1993), the triangles are corrected data for planetary nebulae from 
Leisy \& Dennefeld (1996), the crosses are "corrected" data for planetary 
nebulae from Costa et al. (2000). Open circles are data for H\,{\sc ii} regions 
from Table \ref{table:smc}. 
}
\label{figure:pne-smc}
   \end{figure}

\begin{figure}
\resizebox{\hsize}{!}{\includegraphics[angle=0]{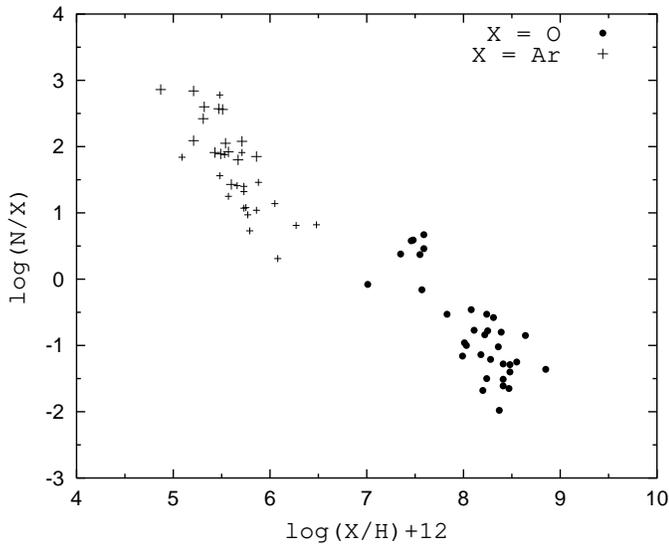}}
\caption{
The N/O -- O/H diagram (points) and N/Ar -- Ar/H diagram (pluses) for 
planetary nebulae in the Small Magellanic Cloud. The original data from 
Leisy \& Dennefeld (1996) and Costa et al. (2000) are shown. 
}
\label{figure:x-nx}
   \end{figure}

The O/H - N/O diagram for H\,{\sc ii} regions and planetary nebulae in the 
Small Magellanic Cloud is shown in Fig.\ref{figure:pne-smc}. 
The points are original data for planetary nebulae from Richer (1993), the 
triangles are the corrected data for planetary nebulae from 
Leisy \& Dennefeld (1996), the crosses are the corrected data for planetary 
nebulae from Costa et al. (2000). The open circles are data for H\,{\sc ii} 
regions from Table \ref{table:smc}. Inspection of Fig.\ref{figure:pne-smc} 
shows that the scatter of oxygen abundances in planetary nebulae of the Small 
Magellanic Cloud is large, more than one order of magnitude. 
An anti-correlation between N/O and O/H is clearly seen. The most evident 
explanation of the scatter is that the oxygen abundance of planetary nebulae 
traces the time of the 
progenitor star formation, the oxygen-poor planetary nebulae progenitors 
being older (Leisy \& Dennefeld 1996, Costa et al. 2000). 

Other possible origins of the oxygen abundance scatter and 
the N/O -- O/H trend 
have also been discussed in the literature. Leisy \& Dennefeld (1996, 2000) 
considered the possibility that the 
oxygen abundance in planetary nebulae does not reflect the initial 
composition of the progenitor star, that it can be destroyed in Type I 
planetary nebulae (He-N rich objects (Peimbert 1978)) as well as produced 
in non-Type I planetary nebulae, 
during the progenitor star life-time.
If this is the major reason for the oxygen abundance scatter 
and the N/O -- O/H 
trend then: {\it i)} the Ar/O ratio would be higher in planetary nebulae with low 
O abundances and {\it ii)} the N/Ar -- Ar/H diagram would differ radically  
from the N/O -- O/H diagram since Ar is not affected by the evolution of  
intermediate-mass stars. Milingo, Henry, \& Kwitter (2002) have undertaken 
a large spectroscopic survey of Galactic planetary nebulae with the goal of 
providing a homogeneous spectroscopic database, as well as a set of consistently 
determined abundances, especially for oxygen, sulfur, chlorine, and argon.
Their data (their Fig.2) do not show a tendency for Ar/O to decrease with O/H.
They found an average Ar/O = 0.0051 
$\pm$ 0.0020 for planetary nebulae, in agreement with 
Ar/O = 0.0055 derived by Izotov \& Thuan (1999) for H\,{\sc ii} regions in 
blue compact dwarf galaxies. 

Fig.\ref{figure:x-nx} shows the N/O -- O/H diagram (points) together 
with the N/Ar -- Ar/H diagram (pluses) for planetary nebulae in the Small 
Magellanic Cloud. The original data from Leisy \& Dennefeld (1996) and Costa 
et al. (2000) are shown. There is both a scatter in argon abundances and 
an anti-correlation between N/Ar and Ar/H. 
Given that Ar/O is independent of O/H 
and that the N/Ar -- Ar/H diagram is similar to the N/O -- O/H diagram, 
one can conclude that possible oxygen production and/or 
destruction during the progenitor star life-time do not dramatically 
change the oxygen abundance of the progenitor star, and oxygen can be used as 
a metallicity tracer, at least in the first approximation. 

The bulk of planetary nebulae in the Small Magellanic Cloud has oxygen abundances 
compatible with those in H\,{\sc ii} regions (Fig.\ref{figure:pne-smc}), 
a fact already 
mentioned by different authors. Based on their model for the chemical evolution 
of the Small Magellanic Cloud, Costa et al. (2000) have concluded that the 
progenitors of these planetary nebulae have been formed 
within the last 1 -- 2 Gyr. 
They also estimated an age of about 12 Gyr for the progenitor of the most 
oxygen-poor planetary nebula in the Small Magellanic Cloud.  

Thus, the oxygen abundances measured in planetary nebulae 
reflect the initial abundance of the progenitor stars while the nitrogen 
abundances are modified by the progenitors' evolution. 
If the initial nitrogen abundance of the progenitor star is known, then the 
difference between the amount of nitrogen in the ejected matter (i.e in 
the planetary 
nebula) and the initial amount of nitrogen in this matter is the 
amount of nitrogen manufactured by the star. The initial nitrogen abundance 
in the progenitor star can be estimated in the following way. Examination of 
Table \ref{table:smc} shows that the variation in the 
N/O ratio from H\,{\sc ii} 
region to H\,{\sc ii} region in the Small Magellanic Cloud is small
within the errors in abundance determinations. Thus we can adopt 
 this N/O ratio 
corresponds as the "equilibrium" ratio in the interstellar medium of 
the Small Magellanic Cloud at the present-day epoch. It is unlikely that 
this value is increased or decreased by the same amount of  
local self-enrichment in nitrogen or oxygen in all HII regions. 
As shown above, the equilibrium N/O ratio in the interstellar medium 
of a galaxy can be increased by the contribution of low-mass stars to the 
nitrogen production. The N/O ratio in the H\,{\sc ii} regions 
of the Small Magellanic Cloud is very close to that in the H\,{\sc ii} regions 
of the most oxygen-poor blue compact galaxies, Fig.\ref{figure:noz-irr}, i.e.  
the present-day equilibrium N/O ratio in the interstellar medium of the Small 
Magellanic Cloud corresponds to the minimum value of the 
N/O ratio. It has not changed up to now, and therefore can be 
adopted as the initial N/O ratio for the progenitor star of any age. Thus, 
the initial nitrogen abundance in the progenitor star can be estimated from 
the oxygen abundance measured in the planetary nebula 
(since this oxygen abundance reflects the initial oxygen abundance of the 
progenitor star) and from the N/O ratio measured in H\,{\sc ii} regions. 
The 
amount of nitrogen manufactured in the star can then be estimated
by subtracting the initial nitrogen abundance from the 
amount of nitrogen observed in the planetary nebula. 
The contribution of low mass stars 
to the nitrogen production can then be compared to that of the intermediate 
mass stars, as shown below.

Galactic Type I (nitrogen-rich) planetary nebulae are believed to correspond 
to the high mass end of the stellar progenitors of planetary nebulae 
(Torres-Peimbert \& Peimbert 1997, and references therein). The planetary nebulae 
in the Small Magellanic Cloud do not appear to fit in this scheme. 
Fig.\ref{figure:pne-smc} shows that nitrogen abundances in oxygen-poor 
planetary nebulae are higher by a factor 3-5 than that in oxygen-rich 
ones. Since the oxygen-poor planetary nebulae in the Small 
Magellanic Cloud seem to be formed from older and therefore less massive stars, then Type I planetary nebulae progenitors can be less 
massive than expected.  Leisy \& Dennefeld (1996) have concluded that, with 
their definition based only on N and He abundances, Type I planetary nebulae 
do not form an homogeneous class.

\begin{figure*}
\resizebox{\hsize}{!}{\includegraphics[angle=0]{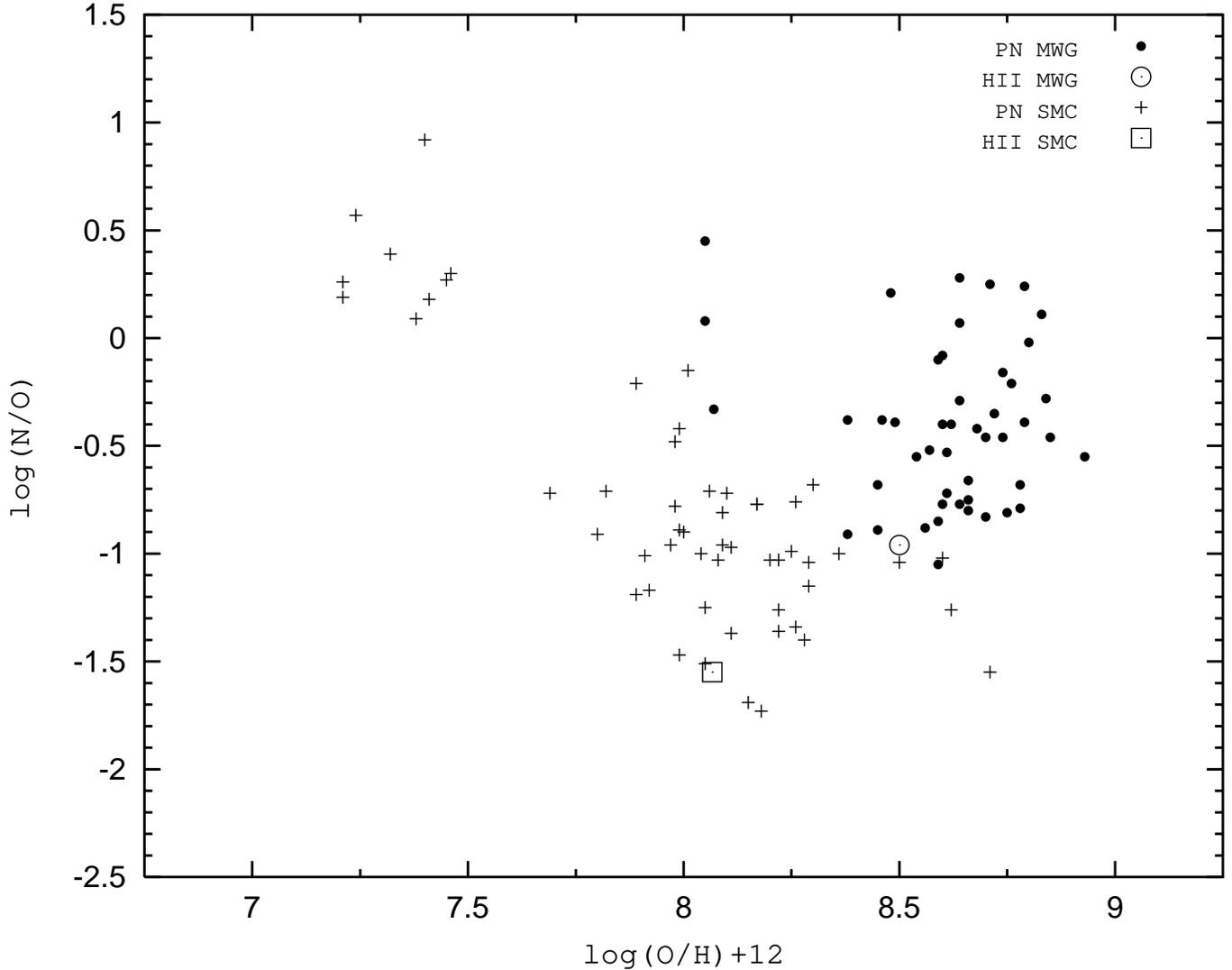}}
\caption{
The O/H - N/O diagram for the planetary nebulae in the Milky Way  
with galactocentric distances between 7.5 and 9.5 kpc (points) and in the 
Small Magellanic Cloud (pluses). The open circle corresponds to the N/O and 
O/H at the solar galactocentric distance (8.5 kpc) as 
traced by H\,{\sc ii} 
regions. The open square corresponds to the average N/O and O/H values in the 
H\,{\sc ii} regions of the Small Magellanic Cloud.
}
\label{figure:pne-mwg}
   \end{figure*}

Let us consider the Milky Way planetary nebulae. In contrast to the Small 
Magellanic Cloud, the Milky Way shows both an oxygen 
and nitrogen radial abundance gradient. 
In order to exclude initial chemical composition differences of the 
planetary nebula progenitors caused by the radial abundance gradient, only 
planetary nebulae in the solar vicinity, with galactocentric distances between 
7.5 and 9.5 kpc, will be considered. Abundances in large samples of 
Galactic planetary nebulae have been 
determined by Aller \& Czyzak (1983), Aller \& 
Keyes (1987), K\"oppen, Acker, \& Stenholm (1991), and Kingsburgh \& Barlow 
(1994). The positions of the Galactic solar vicinity planetary nebulae  
in the N/O -- O/H diagram are shown in Fig.\ref{figure:pne-mwg} by points. 
The galactocentric 
distances were obtained using the distances to planetary nebulae from Cahn 
et al. (1992). When the distance is not given by Cahn 
et al., it is taken from Kingsburgh \& English (1992) or from Maciel (1984). 
The open circle in Fig.\ref{figure:pne-mwg} corresponds to the 
oxygen and nitrogen abundances at the solar galactocentric distance as traced 
by the H\,{\sc ii} regions (Pilyugin, Ferrini, \& Shkvarun, 2002). 
The planetary nebulae in the Small Magellanic Cloud are shown by pluses in 
Fig.\ref{figure:pne-mwg}, the open square corresponds to the average N/O and 
O/H values in the H\,{\sc ii} regions of the Small Magellanic Cloud. 

Inspection of Fig.\ref{figure:pne-mwg} shows that the bulk of Galactic 
planetary nebulae have oxygen abundances in a relatively narrow range. 
These planetary nebulae are believed to be relatively "young", i.e. they 
correspond to the intermediate-mass stellar progenitors of planetary nebulae 
The difference between nitrogen abundances in these planetary nebulae and in 
H\,{\sc ii} regions (this difference reflects the contribution of freshly 
produced nitrogen) is higher than the corresponding value for the 
"young" planetary nebulae of the Small Magellanic Cloud. It is not surprising 
and confirms the widely accepted point of view that the increase of the lower 
envelope of N/O with O/H, beginning at roughly at 12 + log(O/H) = 8.3, is 
explained by the fact that the amount of nitrogen manufactured in the
intermediate-mass stars increases with metallicity.

A few Galactic planetary nebulae have low oxygen abundances. If the oxygen 
abundance of planetary nebula traces the time of the progenitor star formation 
then the oxygen-poor Galactic planetary nebulae progenitors are old low-mass 
stars. However, the possibility that the oxygen-poor 
Galactic planetary nebulae were formed from matter captured by our Galaxy 
from the Small Magellanic Cloud cannot be excluded. The Magellanic Stream is an excellent example of such a present-day capture. 
In this case the oxygen-poor Galactic planetary nebulae are "pseudo-old", 
i.e. they can also correspond to the intermediate-mass stellar progenitors of 
planetary nebulae like the oxygen-rich planetary nebulae. 
Although such a possibility cannot be excluded, for further discussion,
we shall  
consider that the oxygen-poor Galactic planetary nebulae are really "old", 
i.e their progenitors are old low-mass stars. 
 
Comparison between the old planetary nebulae of the Small Magellanic 
Cloud and the Milky Way Galaxy shows that N/O in old planetary 
nebulae slightly decreases with metallicity in contrast to the young planetary 
nebulae which show clearly the tendency for N/O to increase with O/H, 
Fig.\ref{figure:pne-mwg}.  This suggests that the contribution of low-mass 
stars to the nitrogen production decreases with 
metallicity as compared to the contribution of intermediate-mass stars. 
Data on old planetary nebulae in the Milky Way Galaxy are few in 
number, therefore more and better data are needed to verify this suggestion.

The average nitrogen abundance in the old planetary nebulae of the Small 
Magellanic Cloud is around 40-60 N atoms per million H atoms, 
Fig.\ref{figure:pne-smc}. Unfortunately, young planetary nebulae with comparably 
low oxygen abundances (7.3 $<$ 12+log(O/H) $<$ 7.5) are not available,  
therefore a direct comparison of the contributions of low-mass and 
intermediate-mass stars to the nitrogen production for low 
metallicities is  not possible. 
However, if we adopt the usual assumption that the amount of nitrogen manufactured 
by intermediate-mass stars is independent of metallicity in the low 
metallicity range, then the young planetary 
nebulae with oxygen abundances around 12+log(O/H) = 8.1 can be used for 
such a 
comparison. The average nitrogen abundance in the young planetary nebulae of 
the Small Magellanic Cloud is around 10-15 N atoms per million H atoms 
(Fig.\ref{figure:pne-smc}). At the same time, the fraction of the material 
returned into the interstellar medium by the intermediate-mass stars 
(2M$_{\odot}$ $\div$ 8M$_{\odot}$) of a single stellar population is 
comparable with the fraction returned by the low-mass stars ($<$ 2M$_{\odot}$). 
Thus, in the low metallicity range, the contribution of low-mass stars to the 
nitrogen production appears to be rather significant and may even exceed the 
contribution of the intermediate-mass stars. 
The comparison between young planetary nebulae in the Milky Way Galaxy and 
in the Small Magellanic Cloud shows that 
there is a tendency for an increase of the amount of nitrogen 
manufactured in the intermediate-mass stars with metallicity. 
If this tendency holds also in the very low metallicity range,
 then the contribution of low-mass 
stars to the nitrogen production exceeds significantly 
the contribution of the intermediate-mass stars at very low metallicities. 

In summary, consideration of planetary nebulae in the Small Magellanic Cloud 
and the Milky Way Galaxy has shown that the contribution of low-mass stars to the 
nitrogen production is appreciable, and that at low metallicities the 
contribution of low-mass stars can exceed the contribution of the 
intermediate-mass stars. This confirms our conclusion, reached from the 
consideration of the N/O -- O/H diagram for spiral galaxies of different 
morphological types, that there is a long-time-delayed contribution to 
the nitrogen production. 

\section{Discussion}

The present and previous studies suggest that at low metallicities the 
nitrogen production can take place {\it i)} in short-lived massive stars with 
no time delay between the release of nitrogen and oxygen, {\it ii)}  in 
intermediate-mass stars with a moderate time delay of a few hundred Myr between 
the release of nitrogen and oxygen, {\it iii)} in low-mass stars with a 
long time delay of a several Gyr between the release of nitrogen and oxygen. 
It is difficult to differentiate the contributions of massive and 
intermediate-mass stars to the enrichment of interstellar medium in nitrogen on 
the base of available observational data. An attempt to directly estimate the 
nitrogen production by massive stars from observational data of self-enriched 
H\,{\sc ii} regions has not resulted in a firm conclusion.
Then the lowest observed N/O value in low-metallicity H\,{\sc ii} regions 
(where the contribution of the massive stars from current star formation burst 
to the gas enrichment in heavy elements could be dominant) 
can be used as an estimate of the 
nitrogen production by massive stars. The delayed-release hypothesis, i.e. 
the moderate-time-delayed production of nitrogen by intermediate mass 
stars, predicts that N/O 
drops while O/H increases as massive stars begin to die and eject oxygen into 
the interstellar medium. At the point when all the massive stars have died, 
the N/O value is minimum. In the case of a strong star formation burst in 
low-metallicity environment, the 
minimum N/O value corresponds approximately (or exactly in the case of the very 
first star formation burst) to the ratio of N to O yields in massive stars. 

If the conclusion of Izotov \& Thuan (1999) -- that the nitrogen observed in 
low-metallicity (12+log(O/H) $<$ 7.6) blue compact galaxies has been produced 
by massive stars only -- is correct, then the N/O ratio in these galaxies 
(log(N/O) = --1.6) is a lower limit for the global N/O ratio.
In that case, there is little room left for the 
nitrogen production by intermediate-mass stars. However, the N/O 
ratio in the H\,{\sc ii} regions of the Small Magellanic Cloud 
is very close to that in the H\,{\sc ii} regions of the most oxygen-poor blue 
compact galaxies. The bulk of the intermediate-mass stars of the Small
Magellanic Cloud have certainly had enough time to evolve (Pagel \& Tautvai\u{s}iene
1998, Mighell et al. 1998, Rich et al. 2000). If the 
value of log(N/O) = --1.6 corresponds to the element production by massive stars,
then the contribution of intermediate-mass stars to the N/O ratio in the Small 
Magellanic Cloud (log(N/O) = --1.55, Table \ref{table:smc}) 
is negligibly small. 
It is worth noting that while the evolutionary intermediate-mass 
star models predict usually a significant nitrogen production (Renzini \& Voli
1981, van den Hoek \& Groenewegen 1997, Marigo 1996,1998,2001), the non-rotating 
models of Meynet \& Maeder (2002a,b) show very small nitrogen yields by  
intermediate-mass stars in low-metallicity environment. We do not consider the 
very small nitrogen yields by non-rotating intermediate-mass stars of Meynet 
\& Maeder as undisputed argument in favor of negligible small contribution of 
intermediate-mass stars to the enrichment of interstellar gas in nitrogen in 
low-metallicity environment, we only would like to note that such possibility 
cannot be excluded on the base of the existing models.

A crucial argument in favor of a significant production of nitrogen by 
intermediate-mass stars would be the existence of systems with very low N/O 
ratios. Inspection of Fig.\ref{figure:noz-irr} shows that two objects from the 
list of Kobulnicky \& Skillman (1998), CG 1116+51 and Tol 65, have values of 
log N/O appreciably lower than --1.6. However, Izotov et al. (2001) have 
recently redetermined the abundances for Tol 65 (Izotov et al. 2001) and found  
log(N/O) = --1.642$\pm$0.024 for Tol 65, similar to other low-metallicity blue 
compact galaxies. Van Zee et al. (1997, 1998a) have also found 
some galaxies with very low log(N/O) ($<$ --1.7). Izotov \& Thuan (1999) have 
argued that these low values can be caused by systematic errors. The reality 
of the low nitrogen abundances obtained in damped Ly$\alpha$ systems 
(Pettini et al. 1995, Lu et al. 1998, Pettini et al. 2002, Prochaska et al. 
2002) was questioned by Izotov \& Thuan 
(1999). Therefore we will not repeat the discussion of these objects here.
Recall that the low-metallicity range of the observed N/O -- O/H diagram 
for the H\,{\sc ii} regions of irregular and blue compact galaxies 
has appreciably changed over the last decade. Can one exclude the possibility 
that the disagreement between the observed N/O -- O/H diagram 
for the H\,{\sc ii} regions of irregular and blue compact galaxies 
and the observed N/O -- O/H diagram for the damped Ly$\alpha$ systems disappear 
in the near future?

Contini et al. (2002) have discussed the chemical properties of a sample of 
UV-selected galaxies, thought to be in a special stage in 
their evolution, following a powerful starburst. 
Several galaxies in the list of Contini et al. (2002) have values of log(N/O) 
lower than --1.6. Unfortunately, those objects fall 
in or close to the transition region of the R$_{23}$ -- O/H diagram, causing 
a large uncertainty in their abundance determination. 
The oxygen abundances in two objects with low N/O ratio 
(\#30 and \#46) reported by Contini et al. are lower than 12+log(O/H)=8.0. 
Therefore these objects are expected to belong to the lower branch of 
the R$_{23}$ -- O/H diagram, and their abundances can be derived through 
the P -- method for the lower branch of the R$_{23}$ -- O/H diagram. 
This results in 12 + log(O/H)$_P$ = 7.74  for object \#30 (as compared to 
12+log(O/H) = 7.86 found by Contini et al. (2002) using the 
calibration of McGaugh 1991) 
and log(N/O)$_P$ = -1.58  (as compared to 
log(N/O) = -1.78 found by Contini et al. 2002). 
The application of the P -- method to the \#46 object 
results in 12 + log(O/H)$_P$ = 
8.07 (12+log(O/H) = 7.99 was found by Contini et al. 2002). Thus this object 
does not belong to the lower branch of the R$_{23}$ -- O/H 
diagram, and its abundances cannot be found through the P -- method. 
It should be noted that the diagnostic line [OIII]$\lambda 4363$ should 
be detectable in low-metallicity objects like \#30 and \#46 in better 
quality spectra.  

Thus, the level of log(N/O) = --1.6 corresponds to the cumulative nitrogen to 
oxygen yields ratio in massive and intermediate-mass stars.
At present time there are no firm arguments against the conclusion of 
Izotov \& Thuan (1999) that the nitrogen in low-metallicity blue compact 
galaxies has been produced by massive stars only and the level of 
log(N/O) = --1.6 is, in fact, 
defined by the nitrogen to oxygen yield ratio in 
massive stars. On the other hand, the constancy of N/O ratios in 
low-metallicity blue compact galaxies can be equally well reproduced under 
the assumption that a significant part of nitrogen is produced by  
intermediate-mass stars, i.e. the level of log(N/O) = --1.6 is defined by the 
cumulative nitrogen to oxygen yield ratio in massive and intermediate-mass 
stars (Pilyugin 1999). Therefore the adopted contributions of massive and 
intermediate-mass stars to the nitrogen production are based more on belief 
than on proof. 
 
The principal result obtained here is that there is a long-time-delayed 
contribution to the nitrogen production by low-mass stars. 
It is this contribution that 
produces the high log(N/O) $>$ --1.6 in some low-metallicity galaxies. 
Since the N/O ratio in the H\,{\sc ii} regions of the Small 
Magellanic Cloud is close to its lowest level, 
one can conclude that the bulk of 
the low-mass stars of the Small Magellanic Cloud, which are responsible for 
the long-time-delayed contribution to the nitrogen production, 
have not had enough time to evolve. 
Pagel \& Tautvai\u{s}iene (1998) have found that the age-metallicity 
relation for the Small Magellanic Cloud is well reproduced by a model in 
which the majority of the stars are under 4 Gyr old. In this case the low-mass 
stars with lifetimes longer than 4 Gyr are responsible for the long-time-delayed 
contribution to the nitrogen production. Recently Rich et al. (2000) have 
studied the age distribution of star clusters in the Small Magellanic Cloud. 
They have found that the clusters appear to have formed in two brief intervals, 
the oldest 8$\pm$2 Gyr ago and a more recent burst 2$\pm$0.5 Gyr ago.
If the history of cluster formation reflects the total star formation history 
in the Small Magellanic Cloud, then the low-mass stars with lifetimes longer 
than 8$\pm$2 Gyr are responsible for the long-time-delayed contribution to the 
nitrogen production. 

As noted above, it is now commonly accepted now that the increase of the N/O 
with O/H at higher metallicities is caused by the metallicity-dependent
nitrogen production in intermediate-mass stars in high-metallicity environment. 
However, one comment should be made.
Henry et al. (2000) have fit their model to the N/O versus O/H trend for 
a sample of galaxies, and have found that 
nitrogen demonstrates a steeper dependence on metallicity than expected from 
a simple secondary behaviour. Thurston et al. (1996) have determined the slopes 
of a linear fit to the N/O versus O/H data for individual galaxies in their 
sample, and have found that N/O is varying less steeply with metallicity, with
slopes between 0.15 and 1.14, as compared with the 1.0 expected from a 
simple secondary behaviour. The slopes of a linear fit to the N/O versus O/H 
data for individual galaxies in our sample show also a large spread 
(Fig.\ref{figure:noz-type}), 
but about 2/3 of the galaxies show slopes 
less 1.0. The shape of the N/O trend predicted by the model of Henry et al. 
(2000) corresponds well to the shape of the lower 
envelope exhibited by the data (see their Fig.2b). However, 
close examination of our 
Fig.\ref{figure:noz-type} clearly shows that the N/O vs. O/H trends for 
individual galaxies do not correspond well to the shape 
of the lower envelope for all galaxies. 
Thus, the use of a  mix of data including 
galaxies with different star formation 
histories -- and consequently, with different histories of heavy 
element enrichment -- can result in misleading conclusions
 on the behaviour of N/O with O/H.

The value of N/O for a given O/H and the behaviour of N/O with O/H across the
galaxy contain important information about the history of enrichment of 
that galaxy in heavy elements (and consequently, about the star formation 
history). This information is necessary for the construction of 
reliable models for the chemical evolution of galaxies. 
In turn, the construction of detailed 
models for galaxies of different morphological types can help to define more 
precisely the sites of nitrogen production. The construction of 
models for the chemical evolution of galaxies of different morphological 
types will be carried out in future studies. 

\section{Conclusions}

The problem of the origin of nitrogen has been considered within the framework 
of an empirical approach. The oxygen abundances and nitrogen to oxygen 
abundances ratios have been derived in H\,{\sc ii} regions of a number of spiral 
galaxies using the recently suggested P -- method for more than six hundred 
published spectra. The N/O -- O/H diagram for H\,{\sc ii} regions in 
irregular and spiral galaxies has then been constructed. 

It has been found that the level of log(N/O) = --1.6 corresponds to the cumulative 
nitrogen and oxygen productions by massive and intermediate-mass stars.
The constancy of N/O ratios in low-metallicity galaxies can be as well 
reproduced 
under the assumption that massive stars make a dominant contribution 
to the nitrogen production in low-metallicity environments, as under the 
assumption that a significant part of nitrogen is produced by  
intermediate-mass stars. The available data do not allow to decide between 
these two possibilities. 

It has been found that the N/O values in H\,{\sc ii} regions of spiral galaxies 
of early morphological types are higher than those in H\,{\sc ii} 
regions with the same metallicity in spiral galaxies of late morphological 
types. This suggests that there is a long-time-delayed contribution to the nitrogen 
production. The scatter in N/O values at a given O/H can be naturally explained 
by differences in star formation histories in galaxies. The 
low-metallicity dwarf galaxies with  
N/O close to the level of 
log(N/O) = --1.6 
do not contain an appreciable amount of old stars. The 
long-time-delayed contribution to the nitrogen production by old low-mass stars 
is responsible for the high log(N/O) $>$ --1.6 in some low-metallicity galaxies. 

The N/O ratio of a galaxy is an indicator of the time that has elapsed since 
the bulk of star formation occurred, or in other words of the nominal "age" of the 
galaxy as suggested by Edmunds \& Pagel (1978). 

Consideration of the planetary nebulae in the Small Magellanic Cloud and 
the Milky Way Galaxy suggests that the contribution of low-mass stars to the 
nitrogen production is appreciable, and at low metallicities the 
contribution of low-mass stars can exceed the contribution of the 
intermediate-mass stars. This confirms our conclusion, reached from the 
consideration of the N/O -- O/H diagram for spiral galaxies of different 
morphological types, that there is a long-time-delayed contribution to 
the nitrogen production.

\begin{acknowledgements}
We thank Y.~Izotov and J.~K\"oppen for useful discussions. We thank the 
referee, G.Meynet, for constructive comments. 
This study was partly supported (L.S.P.) by the Joint Research Project between 
Eastern Europe and Switzerland (SCOPE) No. 7UKPJ62178, the NATO grant 
PST.CLG.976036, and the Italian national grant delivered by the MURST. 
\end{acknowledgements}

\end{document}